\documentclass[conference]{IEEEtran}
\usepackage{tikz}
\usepackage{footnote}
\usepackage{marginnote}
\usepackage{microtype}
\usepackage{cite}
\usepackage[hyphens]{url}
\usepackage{relsize}
\usepackage[numbers]{natbib}
\usetikzlibrary{external, fit, positioning, backgrounds, shapes.geometric, positioning, calc, shapes, matrix, shapes.gates.logic.US}
\usepackage{relsize}
\usepackage[ruled]{algorithm2e}
\usepackage[noend]{algpseudocode}
\usepackage[caption=false,font=footnotesize]{subfig}
\usepackage{booktabs}
\usepackage{tablefootnote}

\newcounter{figurecounter}

\AtBeginDocument{%
  \providecommand\BibTeX{{%
    \normalfont B\kern-0.5em{\scshape i\kern-0.25em b}\kern-0.8em\TeX}}}

\newcommand{\ttbreakbefore}[1]{%
  \hskip 0pt plus 1fil\relax
  \penalty0
  \hskip 0pt plus -1fil\relax
  \texttt{#1}%
}

\renewcommand{\baselinestretch}{0.95}
\begin{document}
\title{HPIPE: Heterogeneous Layer-Pipelined and Sparse-Aware CNN Inference for FPGAs}

\author{\IEEEauthorblockN{Mathew Hall\IEEEauthorrefmark{1} and
Vaughn Betz\IEEEauthorrefmark{2}}
\IEEEauthorblockA{Department of Electrical and Computer Engineering, University of Toronto\\
Toronto, Canada\\
Email: \IEEEauthorrefmark{1}mathew.hall@mail.utoronto.ca,
\IEEEauthorrefmark{2}vaughn@ece.utoronto.ca}}

\maketitle
\thispagestyle{plain}
\pagestyle{plain}
\begin{abstract}
We present both a novel Convolutional Neural Network (CNN) accelerator architecture and a network compiler for FPGAs that outperforms all prior work. Instead of having generic processing elements that together process one layer at a time, our network compiler statically partitions available device resources and builds custom-tailored hardware for each layer of a CNN.  By building hardware for each layer we can pack our controllers into fewer lookup tables and use dedicated routing. These efficiencies enable our accelerator to utilize 2x the DSPs and operate at more than 2x the frequency of prior work on sparse CNN acceleration on FPGAs. We evaluate the performance of our architecture on both sparse Resnet-50 and dense MobileNet Imagenet classifiers on a Stratix 10 2800 FPGA. We find that the sparse Resnet-50 model has throughput at a batch size of 1 of 4550 images/s, which is nearly 4x the throughput of NVIDIA's fastest machine learning targeted GPU, the V100, and outperforms all prior work on FPGAs.
\end{abstract}
\definecolor{c_add}{HTML}{B5DDD1}
\definecolor{c_bias}{HTML}{D7E7A9}
\definecolor{c_pool}{HTML}{D3C0F9}
\definecolor{c_relu}{HTML}{F99A9C}
\definecolor{c_conv}{HTML}{FDBCCF}
\definecolor{c_placeholder}{HTML}{F3FCA9}
\definecolor{cbuffered}{HTML}{201060}
\definecolor{c_new_line}{HTML}{076678}
\definecolor{c_data}{HTML}{8f3f71}
\definecolor{c_data_valid}{HTML}{427b58}
\definecolor{c_stop_sending}{HTML}{9d0006}
\tikzstyle{buffered}=[draw=cbuffered, very thick]
\colorlet{c_mux}{white!50!c_conv!50!yellow}
\colorlet{c_weights}{white!80!c_conv!60!red}
\pgfdeclarelayer{background}
\pgfdeclarelayer{lines}
\pgfdeclarelayer{control_lines}
\pgfsetlayers{background,control_lines,lines,main}

\tikzset{mul/.style n args={0}{
    path picture={
      \draw[]
        (path picture bounding box.south east) -- (path picture bounding box.north west)
        (path picture bounding box.south west) -- (path picture bounding box.north east);
    }
}}
\tikzset{add/.style n args={0}{
    path picture={
      \draw[]
        (path picture bounding box.south) -- (path picture bounding box.north)
        (path picture bounding box.west) -- (path picture bounding box.east);
    }
}}
\newcommand{\cube}[7]{
    \coordinate (size) at (#1, #2, #3);
    \coordinate (half_size) at ($(size)!0.5!(0,0,0)$);
    \coordinate (middle) at (#4, #5, #6);
    \coordinate (start) at ($(middle)+(half_size)$);
    \draw [-,#7] (start) -- ++(-#1,0,0) -- ++(0,-#2,0) -- ++(#1, 0, 0) -- cycle;
    \draw [-,#7] (start) -- ++(0,0,-#3) -- ++(0,-#2,0) -- ++(0, 0, #3) -- cycle;
    \draw [-,#7] (start) -- ++(0,0,-#3) -- ++(-#1,0,0) -- ++(0, 0, #3) -- cycle;
}
\newcounter{weightcolour}
\newcounter{actcolour}

\newcounter{wavenum}

\setlength{\unitlength}{1cm}
\newcommand*{\clki}{
  \draw (t_cur) -- ++(0,.3) -- ++(.5,0) -- ++(0,-.6) -- ++(.5,0) -- ++(0,.3)
    node[time] (t_cur) {};
}

\newcommand*{\bitvector}[3]{
  \draw[fill=#3] (t_cur) -- ++( .1, .3) -- ++(#2-.2,0) -- ++(.1, -.3)
                         -- ++(-.1,-.3) -- ++(.2-#2,0) -- cycle;
  \path (t_cur) -- node[anchor=mid,yshift=-1pt] {#1} ++(#2,0) node[time] (t_cur) {};
}

\newcommand*{\known}[2]{
    \bitvector{\texttt{#1}}{#2}{white}
}

\newcommand*{\unknown}[2][\texttt{XXX}]{
    \bitvector{#1}{#2}{red!50}
}

\newcommand*{\bit}[2]{
  \draw (t_cur) -- ++(0,.6*#1-.3) -- ++(#2,0) -- ++(0,.3-.6*#1)
    node[time] (t_cur) {};
}

\newcommand*{\unknownbit}[1]{
  \draw[ultra thick,black!50] (t_cur) -- ++(#1,0) node[time] (t_cur) {};
}

\newcommand{\nextwave}[1]{
  \path (0,\value{wavenum}) node[left] {#1} node[time] (t_cur) {};
  \addtocounter{wavenum}{-1}
}

\newcommand{\clk}[2]{
    \nextwave{#1}
    \FPeval{\res}{(\wavewidth+1)/#2}
    \FPeval{\reshalf}{#2/2}
    \foreach \t in {1,2,...,\res}{
        \bit{\reshalf}{1}
        \bit{\reshalf}{0}
    }
}

\newcommand{\dividetime}[2]{
  \draw [draw=none, fill=white] (#1,0.5) .. controls (#1-0.2,-1*#2/3-0.166) .. (#1,-1*#2/2-0.5) .. controls (#1+0.2,-2*#2/3-1.33) and (#1+0.2,-2*#2/3-1.33) .. (#1,-1*#2-0.5) -- (#1+0.1,-1*#2-0.5) .. controls (#1+0.3,-2*#2/3-1.33) and (#1+0.3,-2*#2/3-1.33) .. (#1+0.1,-1*#2/2-0.5) .. controls (#1-0.1,-1*#2/3-0.166) .. (#1+0.1,0.5);
  \draw (#1,0.5) .. controls (#1-0.2,-1*#2/3-0.166) .. (#1,-1*#2/2-0.5) .. controls (#1+0.2,-2*#2/3-1.33) and (#1+0.2,-2*#2/3-1.33) .. (#1,-1*#2-0.5);
  \draw (#1+0.1,-1*#2-0.5) .. controls (#1+0.3,-2*#2/3-1.33) and (#1+0.3,-2*#2/3-1.33) .. (#1+0.1,-1*#2/2-0.5) .. controls (#1-0.1,-1*#2/3-0.166) .. (#1+0.1,0.5);
}

\newcommand{\edgearrow}[5]{
    \coordinate (start) at (#1, -1*#2);
    \coordinate (end) at (#3, -1*#4);
    \coordinate (ystart) at (end|-start);
    \coordinate (yend) at (start|-end);
    \draw [->,#5] (start) .. controls ($(start)!0.6!(ystart)$) .. ($(start)!0.5!(end)$) .. controls ($(end)!0.6!(yend)$) and ($(end)!0.4!(yend)$) .. (end);
}

\newenvironment{wave}[3][clk]{
  \begin{tikzpicture}[draw=black, yscale=.7,xscale=1]
    \tikzstyle{time}=[coordinate]
    \setlength{\unitlength}{1cm}
    \def\wavewidth{#3}
    \setcounter{wavenum}{0}
    \nextwave{#1}
    \foreach \t in {0,1,...,\wavewidth}{
      \clki
    }
}{\end{tikzpicture}}

\section{Introduction}

Recent work has shown that we can maintain accuracy and prune nearly 90\% of the weights from neural networks \cite{efficient_sparse}.  If computation of the pruned parameters can be skipped with custom hardware accelerators, we can potentially realize latency and throughput improvements of up to 10x; however, the pruning process makes the data layout irregular and more challenging to efficiently accelerate.  Recent works have tried to overcome this challenge by taking advantage of the flexible logic, routing, and memories available on FPGAs, but works targeting sparse Convolutional Neural Networks (CNNs) have been limited by low multiplier utilization and some inefficiencies in mapping certain layers to their hardware architecture \cite{efficient_sparse}.

Layers in a neural network each have a different set of properties (e.g. input dimensions, stride, kernel size, etc.) that make it challenging to optimize one type of Processing Element (PE) that is efficient for all of them, and the less regular structure of sparse CNNs heightens this challenge.  For example, a convolutional layer with a kernel size of $7\times 7$ in a neural network with 50 layers imposes a requirement that the PEs must be able to handle kernels that large, even if all of the other layers only use kernels of size $3\times 3$.  During most of the execution time, any additional hardware to support those larger kernels is underutilized.  While not all architectures pay a substantial cost for variable kernel size specifically, they likely pay some penalty for some set of parameters that vary through the network.  Most FPGA and ASIC accelerators use a PE architecture that overprovisions hardware in each PE to handle a variety of layer types.

In this paper we present a novel accelerator architecture that solves these problems by tailoring hardware specifically to each neural network layer and which naturally supports weight sparsity.  We then pipeline across network layers to achieve high throughput and low latency.  Since building custom PEs for each layer of every CNN architecture manually would be intractable, we also developed a network compiler that accepts a TensorFlow graph as input, performs per-layer optimizations, and produces a synthesizable verilog accelerator that implements the network.  We integrate this with a PCIe core and validate network accuracy and accelerator throughput in physical hardware.

The contributions of this paper are as follows:
\begin{itemize}
    \item The novel HPIPE architecture that allows layer-specific optimizations and dramatically reduces the amount of soft logic required to implement 0-weight skipping in convolutional neural networks.
    \item An automated flow that converts TensorFlow network models directly to an optimized HPIPE hardware implementation.
    \item An evaluation of the HPIPE architecture on a sparse ResNet-50 and two variants of MobileNet that both demonstrate the efficiency of our approach and yield higher throughput at batch size 1 than any FPGA or GPU solutions known to the authors.
\end{itemize}

\section{Background and Related Work}


\subsection{Convolutions}
There are three types of 2D convolutions discussed in this work.  The most basic is a 2D convolution that operates on a 3D tensor and outputs a 3D tensor.  The weights have 4 dimensions, $k_h\times k_w \times c_i \times c_o$, where $k_h$ is the height of the weight tensor, $k_w$ the width, $c_i$ the size of the $z$ dimension of the 3D input tensor, and $c_o$ the size of the $z$ dimension of the output 3D tensor.  At each (x,y) point in the input image $c_o$ slices of the kernel with shape $k_h\times k_w \times c_i$ are multiplied element-wise and reduced by summation to produce $c_o$ output points.

The second type of convolution is a pointwise 2D convolution, which is a special case of the standard 2D convolution where $k_h\times k_w=1\times 1$.

The final type is what is called a depthwise convolution.  In contrast to the basic convolution, the kernel is of shape $k_h\times k_w \times c_i \times n$ where $n$ is a channel multiplier that is typically 1.  Also in contrast to the basic 2D convolution, the summation following the element-wise product reduces along only the $x$ and $y$ dimensions, so while there is no $c_o$, the number of channels is preserved (or multiplied by the channel multiplier).

\subsection{Sparsity and Weight Pruning}
Weight pruning is the process of removing (assigning to zero) unimportant weights in a trained neural network.  Recent works have shown that as many as 90\% of the weights can be removed without impacting classification accuracy \cite{efficient_sparse,pruning}.  This offers the opportunity to save memory and memory bandwidth by storing compressed weights, and also the opportunity to skip ineffectual multiplications by pruned (zero) weights.

\subsection{Related Work}
Many accelerators targeting FPGAs have taken advantage of sparsity for Fully Connected (FC) or Long Short Term Memory (LSTM) units \cite{eie,ese,sparse_fc,bank_balanced}.  But fewer have taken advantage of sparsity in convolutional layers.  FC layers and LSTMs have no weight re-use and are memory bound to begin with, so weight pruning provides both a computational and memory bandwidth reduction.  By contrast convolutions share many weights so while we can reduce the computational requirements by pruning, the memory bandwidth reductions are less significant and come with an added cost of less regular computation and lower activation re-use.  The authors are aware of two other attempts to accelerate sparse convolutions on FPGAs. 

In \cite{packed_sparse}, \citeauthor{packed_sparse}  used a novel technique to prune weights and subsequently compress them back into dense weights by combining columns of non-overlapping weights together.  The result was a very efficient accelerator, but the technique only works for pointwise convolutions.  They also implemented a shift unit that implements something similar to a depthwise convolution, but their accelerator can only support one very specific type of neural network. 

In \cite{efficient_sparse}, \citeauthor{efficient_sparse} developed a PE-based sparse CNN accelerator that handles a wider variety of convolutions than \cite{packed_sparse}; however, their performance was limited by a lower frequency than comparable dense accelerators (200MHz), poor mappings of particular layers to their PEs, and a low DSP utilization of only 45\% for an application that is primarily multiplication-bound.

A number of sparse CNN accelerator ASICs have been proposed \cite{cambriconx, scnn, nullhop}.  In this paper we qualitatively assess some of the high level architectural features of SCNN \cite{scnn} and determine they do not translate well to FPGAs.  NullHop \cite{nullhop} also provided FPGA device utilization numbers (they validated their design on an FPGA) which show that they consume 83\% of the logic while using only 6.3\% of the DSPs and run at a frequency of only 60MHz.  The zero-weight skipping architecture of \citeauthor{efficient_sparse} \cite{efficient_sparse} finds a more efficient solution for FPGAs, but their soft logic utilization still limits them to only 45\% DSP utilization.

Our approach differs from prior work in how it leverages the hardware of FPGAs.  While other works have proposed generic PEs that are used for all layers, our work dedicates hardware to every component of an input neural network.  This better leverages the programmable routing and logic resources of FPGAs and enables DSP utilization of 87\% on a sparse Resnet-50 model and 89\% on a dense MobileNet-V1 model.

\section{Architectural Choices}
Convolutions are the most computationally expensive operation in a typical CNN.  Other works have reduced their computational complexity by applying Winograd's minimal filtering algorithm \cite{original_winograd, original_dla, sparse_winograd, numerical_stability_winograd} or a Discrete Fourier Transform (DFT) \cite{fft_cnn}.  While both of these techniques can reduce the number of multiplications required to implement a convolution, they limit our flexibility to perform other optimizations such as precision reductions, and make it more complicated to support a wide variety of convolution configurations (stride, kernel shape, etc.) \cite{numerical_stability_winograd}. In addition to the limits on flexibility, a recent trend in CNN architecture has been to separate convolutions into depthwise and and pointwise convolutions \cite{efficientnet}.  Of these two operations, pointwise convolutions are more computationally expensive, but their complexity cannot be reduced with these transforms.  As a result, we look only at direct convolution methods.

\subsection{Scatter or Gather Convolution}
\begin{figure*}
    \centering
    \subfloat[][\centering Sparse Gather-based Convolution showing full computation of a single output]{\includegraphics[page=1]{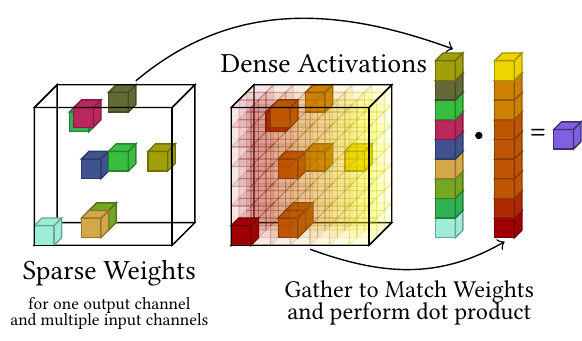}
    \label{fig:gather}}
    \hfil
    \subfloat[][\centering Sparse Scatter-based Convolution showing computation of many partial outputs]{\includegraphics[page=2]{figures/figures.pdf}
    \label{fig:scatter}}
    \caption{Comparison of Scatter and Gather-based Direct Sparse Convolutions}
    \label{fig:scatter_gather_comp}
\end{figure*}

To perform a direct convolution with sparse weights we have the option to either (a) gather the correct activations for a group of weights and multiply and accumulate them in place as shown in Figure \ref{fig:gather}, or (b) multiply all of the weights applicable to a particular input channel with the activations in that input channel and scatter them to a buffer for accumulation, like in \cite{scnn} and as shown in Figure \ref{fig:scatter}.  While the latter may be an efficient choice for an ASIC, on an FPGA we would like to make use of as many hardened resources as possible.  The Stratix 10 DSP blocks include high precision accumulators as well as internal interconnections, called chain-out and chain-in, that allow efficient systolic accumulations within a DSP column, saving power and routing resources.  If we were instead to scatter and accumulate into a buffer outside the DSP column we would require additional soft logic to perform the scatter and addition.  Additionally, accumulation would require both a read and a write every cycle, and streaming completed data out would require another port.  Such a 3-port RAM could be implemented with Stratix 10's quad port RAMs; however, each RAM would then have a maximum width of only 10 bits, which is not suitable for accumulation.  For these reasons we elected to perform a gather-based convolution.

\subsection{Activation Partitioning}
After selecting a convolution method we had to determine a data layout and partitioning scheme to process activations in parallel.  We qualitatively evaluated three different partitioning schemes:
\begin{enumerate}
    \item {\itshape Distribute:} A scheme similar to Intel's DLA \cite{original_dla} that broadcasts activations to PEs from a global buffer and parallelizes multiply accumulate operations across output channels
    \item {\itshape Local Transfer:} A scheme similar to SCNN \cite{scnn} that partitions activations across PEs along the width and height dimensions for a single layer
    \item {\itshape Pipeline:} A scheme that partitions activations in the width and height dimensions across all layers simultaneously
\end{enumerate}

\begin{figure*}[]
    \centering
\vspace{-2em}
    \subfloat[{\itshape Distribute}]{\includegraphics[page=3]{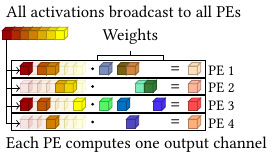}
    \label{fig:dlaarch}}
    \hfil
    \subfloat[{\itshape Local Transfer}]{\includegraphics[page=4]{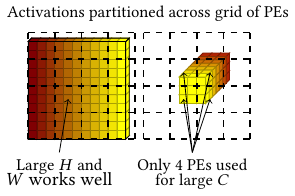}
    \label{fig:scnnarch}}
    \hfil
    \subfloat[{\itshape Pipeline}]{\includegraphics[page=5]{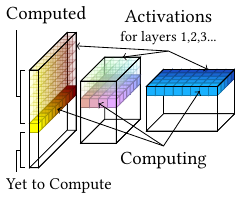}
    \label{fig:pipelinearch}}
    \caption{Overview of Activation Partitioning Architectures Explored}
    \label{fig:archcomp}
\end{figure*}

\subsubsection{Distribute}
Intel's DLA \cite{original_dla} uses a PE architecture that streams and duplicates input features across multiple processing elements that each compute a different output channel.  {\itshape Distribute} works very well for a dense accelerator; however, with a sparse accelerator targeting around 85\% weight sparsity, only 15\% of the activations are used per output channel computation, as illustrated in Figure \ref{fig:dlaarch}.  Since each PE would only use 15\% of the activations broadcast to it, we would either need to have that PE compute multiple output channels serially (which constrains our ability to parallelize the computation) or we would need to increase the activation distribution bandwidth, at a substantial hardware cost, to match the throughput of the DSPs.  Additionally, if we parallelize across output channels then we need each processing element to perform its own address calculations, which are more complicated and expensive for a sparse accelerator.  From this assessment of the {\itshape Distribute} architecture we conclude that a hardware efficient sparse accelerator needs to (a) minimize activation movement since activation re-use is relatively much lower in a sparse accelerator, and (b) share address computations for a large number of output activations computed in parallel.

\begin{table*}[]
    \centering
    \caption{Comparison of Activation Distribution/Partitioning Architectures}
    \label{tab:activation_partitioning}
    \begin{tabular}{lccccc}
      \toprule
                                  & Activation Locality & Address Computation & Shape Flexibility & Weight Bandwidth & Latency \\
      \midrule
        {\itshape Distribute}     & Poor                & Poor            & Good              & Excellent        & Excellent\\
        {\itshape Local Transfer} & Good                & Good            & Poor              & Good             & Excellent\\
        {\itshape Pipeline}       & Excellent           & Excellent       & Excellent         & Poor             & Good\\
    \bottomrule
    \end{tabular}\\[-1em]
\end{table*}

\subsubsection{Local Transfer}
\label{local_transfer}
SCNN \cite{scnn} is a sparse ASIC accelerator that minimizes activation movement by partitioning input activations across tiled PEs in their height and width dimensions for a single layer. In this architecture activations needed by multiple PEs are directly sent to adjacent PEs.  While this solves the activation bandwidth issue we had with {\itshape Distribute},  this partitioning scheme has a PE under-utilization issue since the activations cannot be split across many PEs when the height and width dimensions of the activations shrink. Figure \ref{fig:scnnarch} shows two sets of activations being partitioned across a PE array.  The one with large height and width dimensions works well, but the second one with many channels but small height and width dimensions only utilizes 4 PEs.

\subsubsection{Pipeline}
The last partitioning we considered was to build a fixed pipeline of layers and pass activations directly between the stages.  Figure \ref{fig:pipelinearch} shows how we can have multiple stages each computing a portion of different layers, which we call a partition, in parallel.  Notice that the earlier layers will compute multiple partitions before later stages begin (since they are waiting for data from prior layers).  The primary disadvantage of this architecture is that it requires a tremendous amount of memory bandwidth for weights.  Each of the \emph{Computing} (see Figure \ref{fig:pipelinearch}) partitions require the entire set of weights to finish an output line. To reduce weight memory bandwidth requirements other accelerators typically load a set of weights and multiple input images, then use the weights to complete output activations for multiple inputs.  By contrast, the {\itshape Pipeline} architecture uses all of the weights to complete only a portion of a single input.  It then needs to load all of the weights again to complete the next portion of that input.

\subsection{Comparison}

We have summarized our findings about each of the architectures in Table \ref{tab:activation_partitioning}.  A brief summary of the reasons we assigned different values to each architecture is provided below:
\begin{itemize}
    \item Activation Locality: {\itshape Distribute} requires that activations all come from and are written back to a global buffer on the chip, while {\itshape Local Transfer} transfers them directly to adjacent PEs, and {\itshape Pipeline} passes them directly to a small computation unit where they will next be needed.
    \item Address Computation: {\itshape Distribute} scores poorly here because each relatively small PE would require its own address computation unit.
    \item Shape Flexibility: Section \ref{local_transfer} detailed the way in which {\itshape Local Transfer} has difficulty adapting to changing input size; {\itshape Distribute} also scores slightly lower since PEs will be under-utilized for layers without enough output channels.
    \item  Weight Bandwidth: {\itshape Distribute} and {\itshape Local Transfer} use the weights they load for one layer to compute an entire output, while {\itshape Pipeline} uses all weights in a layer to compute only part of an output, giving it a lower score.
    \item Latency: both {\itshape Distribute} and {\itshape Local Transfer} use all of their multipliers to compute every intermediate activation; {\itshape Pipeline} scores lower here since it requires multiple input images to completely fill the pipeline and take advantage of all of the multipliers.
\end{itemize}

None of the architectures scored perfectly in all categories. The {\itshape Pipeline} architecture performs well across all of our metrics except weight bandwidth, where it performs very poorly.  We can solve this problem by requiring that all weights fit into on-chip storage. That may seem like a substantial disadvantage, but we feel that a combination of (a) the general trend towards reduced parameters in neural networks \cite{efficientnet}, (b) our aggressive parameter reduction through pruning, (c) lower precisions lowering storage requirements, (d) more on-chip memory on newer devices, and (e) Microsoft's approach of connecting multiple FPGAs together to fit an entire network into on-chip storage \cite{brainwave}, make this requirement a fair price to pay for the other substantial benefits {\itshape Pipeline} offers.

With {\itshape Pipeline} we can tailor modules to every single layer in the neural network with optimizations described in Sections \ref{network_compiler} and \ref{detailed_implementation}.  We can also share address computation units across a large number of multipliers in each layer.  Finally, we minimize activation transfer and duplication by passing activations directly from the output of one computational unit to the input buffer of the next.

\section{Network Compiler and Tool Flow}
\label{network_compiler}
\begin{figure}
    \centering
    \vspace{-1.25em}
    \scalebox{0.5}{\includegraphics[page=16]{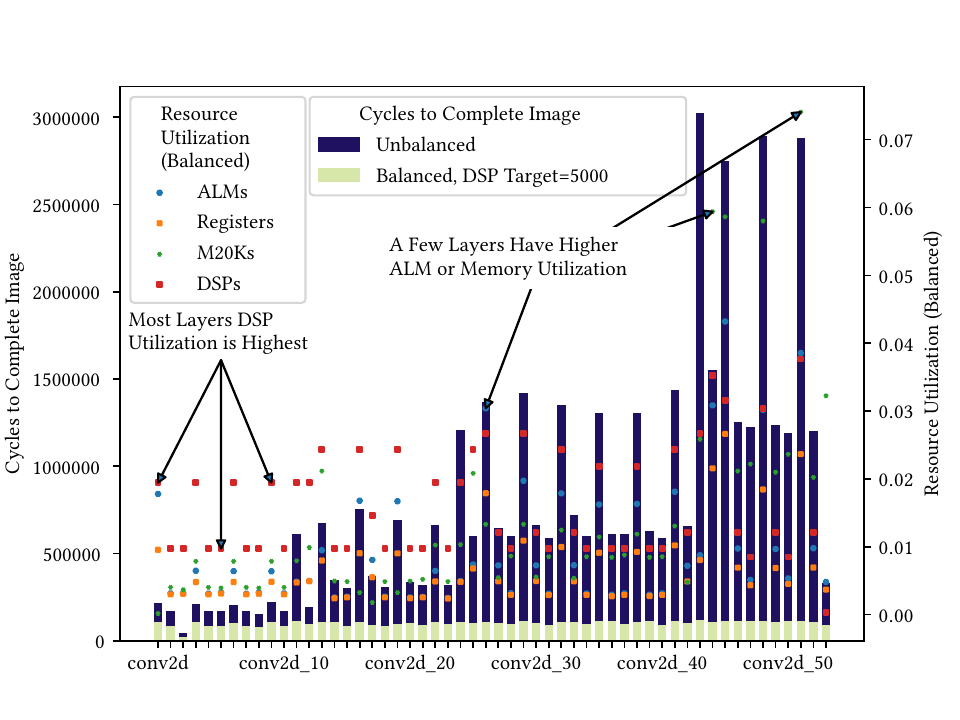}}
    \vspace{-1em}
    \caption{Comparison of Individual Layer Latency Before and After Balancing, and Resource Utilization per Layer as Percentages of the Total Chip Resources}
    \label{fig:cycles_to_complete_image}
    \vspace{-1.5em}
\end{figure}

When mapped to our convolution unit with the default parallelism settings, the layers in ResNet-50 do not all have the same throughput.  Since the slowest stage in a pipeline determines the throughput for the entire pipeline we had to design parameterized hardware that can make use of additional area to balance the throughput of all of the layers.  Figure \ref{fig:cycles_to_complete_image} shows the cycle counts from independent simulations of each of the convolution stages in our accelerator for ResNet-50.  The {\itshape Unbalanced} bars show the cycle counts for stages without parameters optimized for balanced throughput.  The {\itshape Balanced} bars show the cycles after we have recompiled the network to target 5000 Stratix 10 DSP blocks.  The dots show the device utilization for the {\itshape Balanced} accelerator broken down into the fractions of total device ALMs, registers, M20Ks, and DSPs.  Generating these parameters along with the hardware itself and the memory initialization files for the hardware manually would be tedious so we elected to automate this process.

Others have built tools that generate neural network accelerators for FPGAs. \citeauthor{vta} \cite{vta} built a platform independent Intermediate Representation (IR) to which they compile graphs from a number of machine learning frameworks. \citeauthor{legup} \cite{legup} created their own LLVM passes to translate LLVM IR they obtained from TensorFlow's Accelerated Linear Algebra (XLA) back end to LLVM IR that the open source High Level Synthesis (HLS) tool LegUp can accept and convert into Verilog. \citeauthor{dnnweaver} \cite{dnnweaver} built a tool that translates a Caffe protobuf description of a neural network into a series of instructions for their own accelerator, then builds an accelerator tailored to the instructions used by the input network and a particular FPGA. \citet{cloud_dnn} built a tool that maps from Caffe protobufs to HLS templates and builds a full system that can have parts of a network run in software.  While these tools all have something to offer, we elected to implement our own solution.

Like \citeauthor{dnnweaver} \cite{dnnweaver} we wanted to write our own highly optimized Register Transfer Level (RTL) description of a select set of neural network operations to target the highest possible performance; however, our layer-pipelined architecture is too disimilar from theirs to reuse any of their optimized hardware.  The IR from \citeauthor{vta} \cite{vta} would have allowed HPIPE to accept networks from a variety of frameworks; however, mapping from IR to our hardware would have been challenging since we would have needed to infer the type of operation from a control flow graph.  Importing TensorFlow graphs gives us the structure of a wide variety of different neural networks at a level of abstraction that is appropriate for our hardware.

\begin{figure}
    \centering
    \includegraphics[page=8]{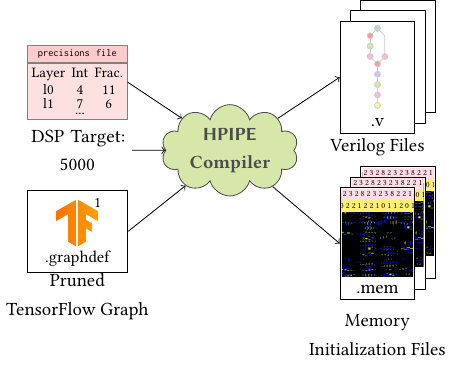}
    \vspace{-1em}
    \caption{HPIPE Compiler Inputs and Outputs}
    \label{fig:graph_compiler}
    \vspace{-1.5em}
\end{figure}

Our compiler accepts a TensorFlow graph, a DSP target that provides it with a coarse estimate of the resources on the target device, and a precision annotations file that allows a user to specify a particular fixed point format independently for each of the operations in the graph.  \cite{wrpn} and \cite{finn_l} found that there is no single optimal precision for neural network acceleration, so the ability to tune the precision of activations and weights independently for each operation has the potential for large efficiency gains. Figure \ref{fig:graph_compiler} shows the overall flow of our compiler from a user perspective.  The output is a directory containing a series of verilog files that implement the input neural network, and a number of memory initialization files that contain compressed weights and auxiliary data that are described in more detail in section \ref{detailed_implementation}.

\footnotetext[0]{1. TensorFlow, the TensorFlow logo and any related marks are trademarks of Google Inc.}

To prepare a TensorFlow graph to be built into an HPIPE accelerator our compiler first attempts to merge all of the batch normalization operations into convolution and bias operations.  While batch normalization must run as an independent operation during training, during inference it simplifies to a multiplication by a constant and an addition with a constant.  Folding batch normalizations into other operations is commonly performed to prepare neural networks for inference, and a utility for performing this operation is included in TensorFlow r1.11; however, this utility only looks to see if the operation can be merged into its immediate neighbours.  We run a series of graph transformations that break batch normalizations into an addition and a multiplication and then swap the execution order of certain operations so that they can be merged with operations that were not initially neighbours.  After these transformations are complete we dump a new TensorFlow graphdef that can be run through TensorFlow to validate that the tranformations did not impact the accuracy of the network.

We allow our compiler to swap batch normalizations with max pool and padding operations, as well as move the multiplication component of the batch normalization after ReLUs. We have run these optimizations on the official TensorFlow ResNet-50 model and found they allowed the compiler to successfully fold all batch normalizations into other operations with no impact to either top 1 or top 5 accuracy on ImageNet \cite{imagenet}.

After merging batch normalizations we merge padding operations into pooling or convolution operations and build a balanced plan for the target DSP count.  To enable balanced throughput our convolution module has an \texttt{n\_channel\_splits} parameter that allows us to unroll the computation of output activations along the input channel dimension.  With an analytic model that estimates the throughput of a convolution operation, given this parameter, we can loop over the slowest operations and increment \texttt{n\_channel\_splits} until we hit the DSP Target.

Initially our model assumed a linear relationship between \ttbreakbefore{n\_channel\_splits} and the throughput of a module.  This proved to be a poor assumption for some layers with a high degree of sparsity due to the distribution of the zeros within that layer.  We rectified this by computing the actual weight partitioning and padding that a later stage of the compiler performs, which improved our estimates to within 1\% of the actual throughput and improved the throughput of the accelerator by 23\%.  As you can see in Figure \ref{fig:cycles_to_complete_image}, our algorithm is able to balance the stages of our 85\% sparse ResNet-50 model such that nearly all of the layers have throughput within 10\% of each other, which results in a throughput improvement of 30x versus the unbalanced design.  The algorithm runtime is only a few seconds.

Once we have come as close to the DSP target as possible without exceeding it, we pass the plan with the computed parameters to our accelerator generator.  This generator iterates over all of the nodes in the optimized TensorFlow graph, instantiating modules for every node with the parameters contained in the plan, then iterates over all of the edges and connects the modules together.  Finally, it dumps a directory containing verilog and memory initialization files that implement the graph for the target FPGA.

\section{Detailed Implementation}
\label{detailed_implementation}
We have implemented and verified modules that can execute the TensorFlow Placeholder, Conv2D, DepthwiseConv2D, MatMul, BiasAdd, MaxPool, Relu, Relu6, Add, and Mean operations.  As described in Section \ref{network_compiler}, we have also implemented a series of graph transformations that allow us to merge all of the BatchNormalization operations in the official TensorFlow ResNet-50 V1 (r1.11), MobileNet-V1, and MobileNet-V2 models into Conv2D and BiasAdd operations.  Our implementations of MaxPool, Conv2D, and DepthwiseConv2D can support any kernel shape and stride and are further parameterized to allow our network compiler to allocate additional device resources to layers that have higher computational requirements.  This section first gives an overview of the the general data flow through the circuit, then it details the internal implementation of the convolution block and provides an overview of the other blocks.

\subsection{Data Flow}
\begin{figure}
    \centering
    \includegraphics[page=13]{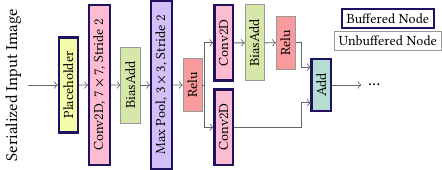}
    \caption{First Layers in ResNet-50 as they Map to our Modules}
    \label{fig:resnet50_modules}
\vspace{-1em}
\end{figure}
Figure \ref{fig:resnet50_modules} shows the first 10 TensorFlow operations from our optimized 85\% weight sparse ResNet-50 V1 model.  Each of these operations is implemented as a module instantiation, and the arrows between them are pipelined wires connecting {\itshape producers} to {\itshape consumers} (for example, in Figure \ref{fig:resnet50_modules} the Placeholder is a {\itshape producer} for the Conv2D it is connected to).  Some of the operations buffer the input data, while others simply process it as it comes in and immediately write it out.  The nodes that have buffers have a limited amount of storage space, so they export a coarse-grained back-pressure signal to all of their producers indicating if there is space in the buffer. The nodes that do not buffer input simply pass the signal from their consumers to their producers.

Each stage in HPIPE processes one line of output data at a time.  We call this an output channel group and it has a shape of $1\times W\times C_o$, where $W$ is the output width and $C_o$ is the number of output channels.  Additionally, all modules contain both a controller and a data path.  The controllers typically compute addresses, load weights and biases, and communicate with their producers and consumers.

\subsection{Convolution and Matrix Vector Multiplication}
\begin{figure}
    \centering
    \includegraphics[page=6]{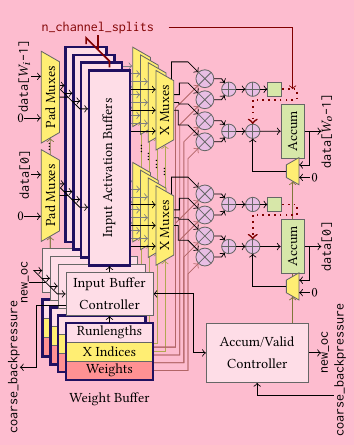}
    \caption{Convolution Module Diagram}
    \label{fig:conv_module}
    \vspace{-2em}
\end{figure}

Figure \ref{fig:conv_module} shows a block diagram of our convolution module (note that we have omitted all of the registers not essential to the functionality of the circuit) with \texttt{n\_channel\_splits} set to 4.  We use this module to implement both convolutions and Matrix-Vector multiplications (a convolution with a kernel shape of $1\times 1\times c_i\times c_o$ with an input shape of $1\times 1\times c_i$ is the same as a matrix-vector multiplication of shape $c_o\times c_i$).  The function of each of the blocks is as follows:
\begin{itemize}
    \item Pad Muxes: For layers with vertical padding the Input Buffer Controller will write zeros into the first $P_t$ lines of the Input Activation Buffers prior to deasserting its coarse backpressure signal (where $P_t$ is the top padding), and do the same for the bottom padding before a new input can be processed.
    \item Input Activation Buffers: Are a series of ring buffers into which the input activations are packed.
    \item Weight Buffer: Stores compressed weights, runlengths that encode the $y$ and $z$ position of a weight as an offset from the position of the previous weight, and x-indices indicate the weight's $x$ position.
    \item Input Buffer Controllers:
    \begin{itemize}
        \item Control when to write padding.
        \item Decode runlengths from the weight buffer into addresses from which activations are read from the Input Activation Buffers.
        \item Store the amount of space remaining in the Input Activation Buffers and assert \texttt{coarse\_backpressure} if there is no longer enough space to write a full line (all channels).
    \end{itemize}
    \item X Muxes: One $k_w$-to-1 mux for each multiplier that allows selection of activations from different $x$ locations.
    \item DSPs:  The circles with Xs are multipliers, and each DSP contains two multipliers, plus two adders (circles with +s) and either an accumulation register or a delay register.  The dotted red lines in Figure \ref{fig:conv_module} from the delay register in one DSP block to the adder in the next DSP block indicates chaining of \texttt{n\_channel\_splits}/2 DSP blocks together.
    \item Accum/Valid Controller: Has a memory containing the number of weight lines per output channel that it loads into a down counter that stops accumulation and asserts the \texttt{new\_oc} signals whenever it reaches zero.
\end{itemize}

At a high level the operation of the block is as follows:
\begin{enumerate}
    \item Data enters from producers through \texttt{data} lines 1 to $W$, and gets written whenever the \texttt{new\_oc} signal is asserted.
    \item Every time the input \texttt{new\_oc} is asserted the address to which the input activatons are written is incremented.
    \item Once the buffer contains $k_h$ full input lines (where $k_h$ is the kernel height for the layer), it begins to read from the weight buffer.
    \item The Input Buffer Controller begins to decode runlengths into addresses and distributes them to the input activation buffers.
    \item The activations loaded by the Input Activation Buffer are shifted by the X Muxes by the amount specified by the corresponding X Index from the Weight Buffer.
    \item The shifted activations are multiplied with the corresponding weight and accumulated.
\end{enumerate}

\begin{figure}
    \centering
    \includegraphics[page=15]{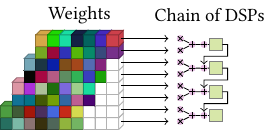}
    \caption{Weight Data Shifting for Systolic Chaining}
    \label{fig:weight_data_shifting}
\vspace{-1.5em}
\end{figure}

At the top of Figure \ref{fig:conv_module} you can see the \texttt{n\_channel\_splits} parameter that we use to configure the throughput of the convolution block.  As shown in the diagram \texttt{n\_channel\_splits} is the number of Weight Buffers, Input Buffer Controllers, Input Activation Buffers, and X Muxes we instantiate.  It also controls the number of multipliers; however, it does not control the number of accumulators.  As we increment \texttt{n\_channel\_splits} and add additional multipliers, we use the chain-out from the DSP to string multiple DSPs together, using only the last one in the chain to perform the accumulation.  Since each additional DSP in the chain introduces one cycle of delay, we shift the Weights, X Indices, and Runlengths for every second channel split as shown in Figure \ref{fig:weight_data_shifting}.  The alternative would have been to build parallel reduction trees in the soft logic, which would have been quite expensive.

\subsection{Other Modules}
The other modules have a similar layout to the Conv2D module, except that BiasAdd and Relu have no buffer, and Placeholder has a FIFO instead of an Input Activation Buffer.  The Add operation has one Input Activation Buffer for each producer module, and the depth of each of these buffers is computed to ensure the amount of buffering on skip paths matches the amount of buffering on the non-skip path.  This ensures that our pipeline does not run into a deadlock where an Add operation is waiting on data from a path that cannot receive data because another path consuming that data has full buffers.


\section{Performance Results}
\label{performance_results}
We compare to three sets of related accelerators.  First, we evaluate our throughput on Resnet-50, a popular and high accuracy but compute-intensive CNN.  The highest performance GPU and FPGA implementations of Resnet-50 are dense, so we compare to these, but our implementation leverages sparsity and achieves higher throughput at a modest accuracy cost.  Second, we compare to a prior sparse-CNN FPGA accelerator and show that we can exploit all the device DSPs, while their higher logic utilization limits the number of DSPs they can use. Third, we compare throughput vs. a GPU and a prior FPGA accelerator on the compute-efficient and \emph{dense} MobileNet CNNs (V1 vs. a GPU and V2 vs. the prior FPGA accelerator).  This comparison shows we outperform these accelerators even without leveraging sparsity and while running at 2x the precision.

\subsection{Highest Throughput Accelerators}
\begin{figure}
    \centering
    \scalebox{0.5}{
    \tikzsetnextfilename{throughputVsLatency}
    \input{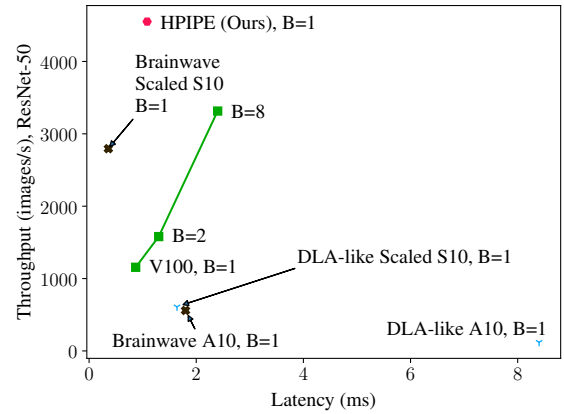}
    \addtocounter{figurecounter}{1}
}
    \caption{Throughput vs. Latency Comparison with V100 \cite{v100}, Brainwave \cite{brainwave}, and DLA-Like \cite{original_dla} Accelerator on Resnet-50, B=$N$ denotes batch size of $N$}
    \label{fig:perf}
\vspace{-1.25em}
\end{figure}

\begin{table*}[h]
    \centering
    \caption{Resource Counts and Utilization for HPIPE on Stratix 10 2800}
    \label{tab:resource_counts}
    \begin{tabular}{l c c c c c c c}
      \toprule
        CNN       & ALMs Needed      & ALMs for Memory & ALM Registers   & Hyper-Registers & M20Ks & DSPs & Frequency\\
    \midrule         
        Resnet-50 & 591,882 (63\%)   & 122,850 (26\%)  & 1,417,297 (37\%)& 372,592         &11,278 (96\%)& 5,022 (87\%) & 580 MHz\\
        MobileNet-V1 & 371,500 (40\%)& 110,950 (24\%)  & 874,713 (23\%)  & 147,671         & 4,283 (37\%) & 5,133 (89\%)  & 430 MHz\\
        MobileNet-V2 & 290,486 (31\%)& 41,550 (9\%)    & 766,604 (21\%)  & 105,810         & 4,512 (38\%) &  2,964 (51\%)  & 390 MHz\\
    \bottomrule
    \end{tabular}
    \\[-2em]
\end{table*}
On Resnet-50 we evaluate against the highest performance machine learning optimized GPU (an NVIDIA V100 with up to 260 trillion operations per second of 8-bit matrix multiply performance) using the optimized numbers reported by NVIDIA \cite{v100}, an academic performance model of Intel's DLA \cite{original_dla} (we refer to this as DLA-Like), and Microsoft's Brainwave \cite{brainwave}. The only available numbers for DLA-Like and Brainwave are from Arria 10 (A10) FPGAs, so we also provide scaled numbers for Stratix 10 (S10). For DLA-Like we scaled them by a compounded 3.4x for the $\sim$2.3x increase in $18\times 18$ multipliers and a 1.5x improvement in frequency.  For Brainwave we use the Peak TFLOPs numbers they provide for S10 and A10 to scale their A10 throughput and latency numbers.  Figure \ref{fig:perf} shows the throughput versus latency of each of these accelerators.  Since the GPU has throughput improvements when it is run with larger mini-batches we have shown the latency throughput trade-off curve and annotated the batch size.  HPIPE has nearly 4x the throughput of the V100 at a batch size of 1. Moving up to a batch size of 8 the V100 has 72\% of the throughput, but with 2.2x the latency, and the requirement that it must batch 8 images.  Similarly, HPIPE outperforms Brainwave and DLA-Like by 1.6x and 7.4x, respectively, even when we compare to scaled numbers that assume throughput scales perfectly with peak TFLOPs.

\begin{table}[]
\begin{center}
    \caption{Performance Optimizations and Impact on Accuracy -- Resnet-50}
    \label{tab:performance_optimizations}
    \begin{tabular}{p{7em} c c c c}
      \toprule
           & V100 \cite{v100} & Brainwave \cite{brainwave} & HPIPE & DLA-Like \\
    \midrule
    Sparsity & 0\% & 0\% & 85\% & 0\%\\
    Winograd & No & No & No & Yes \\
    Precision & 8-Bit & 11-Bit & 16-Bit & 16-Bit\\
    Format & Fixed & Block Float & Fixed & Fixed\\ 
    Top-1 Accuracy & 74.93\% &  $\sim 76\%$\footnotemark[2]     & 71.9\% & --\\
    \bottomrule\vspace{-1em}
    \end{tabular}
\end{center}
    \footnotesize{\footnotemark[2] \citet{brainwave} provides only a statement that accuracy degradation is negligible}\vspace{-2em}
\end{table}

Each of the devices are running variants of ResNet-50; however, there are slight differences in each of the models summarized in Table \ref{tab:performance_optimizations}.  We collected our accuracy by running the 50,000 image ImageNet validation set on actual hardware, using PCIe to transfer data to and from the accelerator. Our top-1 and top-5 accuracies of 71.9\% and 90.8\% match the input TensorFlow model when it is run on a GPU. This is lower than typical ResNet-50 accuracy of top-1 and top-5 of 76.0\% and 93.0\%\cite{resnet}, respectively; however, we believe we could increase this with a different pruning technique that does not restrict us to the same sparsity in each layer.

\subsection{Sparse CNN on FPGA}
\citeauthor{efficient_sparse} \cite{efficient_sparse} do not provide latency or throughput numbers for their accelerator, so we can only compare resource utilization. Table \ref{tab:resource_utilization} shows that our frequency is nearly 3x theirs, and our DSP utilization is nearly double theirs.  They use a smaller FPGA, but we expect that if they scaled up to a larger FPGA, their DSP-to-logic utilization ratio would remain roughly the same and their accelerator would still be unable to take advantage of all of the available multipliers.

\subsection{Dense MobileNet-V1 and MobileNet-V2}

Table \ref{tab:mobilenet_throughput} shows a comparison of HPIPE to the V100 GPU running MobileNet-V1 and a comparison of HPIPE to the FPGA accelerator from \citet{xilinx_layer_pipe} running MobileNet-V2.  NVIDIA does not report accuracy for their implementation of MobileNet-V1.  In the comparison against the GPU we are behind in latency by 0.43ms, but we demonstrate higher throughput despite running at 2x the precision and not leveraging the sparse acceleration capabilities of our accelerator.

While the V100 and S10 2800 are both the largest monolithic chips sold by NVIDIA and Intel, the Zynq ZU9 is not, so we must normalize the results in some way to compare to \citet{xilinx_layer_pipe}.  We cannot simply divide throughput by DSP count either, because one Xilinx Ultrascale+ DSP slice is a single 27x18 multiplier while the Intel S10 DSP block is two 18x18 multipliers.  To ensure our comparison is rigorous we will divide our throughput by the number of 18x18 multipliers we use and divide their throughput by the number of 27x18 multipliers they use (despite the 27x18 multipliers taking more area).  Doing this yields throughput per multiplier 1.95x higher for HPIPE than \citet{xilinx_layer_pipe} even though they use half our precision.  We believe this performance derives from a combination of a) our higher frequency of 390MHz vs. 333MHz and b) under-utilization of their DSP blocks due to imperfect mapping of some layers onto their PEs.

Unlike MobileNet-V1 and Resnet-50, for MobileNet-V2 we were unable to fully utilize the S10 2800 DSPs.  The current version of HPIPE only unrolls the input channel dimension and we ran out of input channels to unroll.  In the future we could update HPIPE to unroll output channels to extract additional parallelism.  As it is, our DSP utilization is still higher than our soft logic or M20K utilization, and we could fit on an S10 1650 and utilize 94\% of the DSPs.

\subsection{Resource Utilization and Physical Optimization}
Table \ref{tab:resource_counts} provides a more detailed breakdown of our resource utilization for the three models we evaluate. Resnet-50 is memory bound, using 96\% of the M20Ks, and both MobileNet models run out of parallelism to extract prior to running out of any device resources.  The frequencies for each of the HPIPE CNNs are higher than any of the FPGA-based accelerators to which we compare, though the two MobileNet accelerators have a lower frequency than the Resnet accelerator.  Our compiler adds additional pipeline stages to control and data signals based on fanout count and some estimates of the area over which these fanouts span.  These are heuristic estimates that were mostly tuned on Resnet.  Future optimization could yield higher frequencies for both MobileNet models.

\begin{table}[]
\setlength{\tabcolsep}{4pt}
\renewcommand{\arraystretch}{1.2}
    \centering
    \caption{Dense MobileNet Accelerator Comparison}
    \label{tab:mobilenet_throughput}
    \begin{tabular}{p{5em} c c c c}
      \toprule
         & \citet{xilinx_layer_pipe} & HPIPE (Ours) & V100 \cite{v100} & HPIPE (Ours)\\
    \midrule    
        Device & Zynq ZU9 & S10 2800 & V100 & S10 2800\\
        DSPs Used   & 2,070     & 2,964   & -- & 5,022 \\
        MobileNet Version & V2 & V2 & V1 & V1 \\
        Precision \footnotesize{($Bits$)} & 8 & 16 & 8 & 16\\
        Throughput \footnotesize{($B$=$1$, $im/s$)} & 810 & 4,539 & 4,605 & 5,157\\
        Latency \footnotesize{($B$=$1$, $ms$)} & -- & 1.1 & 0.22 & 0.65 \\
        Top-1 \ \ \ \ \ \  Accuracy & 68.1\% & 71.9\% & -- & 71.7\%\\
    \bottomrule
    \end{tabular}
    \vspace{-1em}
\end{table}

\begin{table}[]
    \centering
    \caption{Resource Utilization Comparison with Prior Sparse CNN Accelerator on Resnet-50}
    \label{tab:resource_utilization}
    \begin{tabular}{l c c}
      \toprule
         & \citet{efficient_sparse} & HPIPE (Ours)\\
    \midrule    
        Device & Xilinx Zynq ZCU102 & Intel Stratix 10 2800 \\
        Frequency (MHz) & 200 & 580\\
        Logic Utilization & 92\% & 63\%\\
        DSP Utilization & 45\% & 87\%\\
        BRAM Utilization & 48\% & 96\%\\
    \bottomrule
    \end{tabular}
    \vspace{-1.5em}
\end{table}

\section{Future Work}
\label{future_work}
HPIPE affords significant flexibility to optimize both the precision and the sparsity of each layer.  In the current paper we ran all of our experiments with a 16-bit fixed point precision.  With future Agilex FPGAs including 2x performance for 8-bit vector dot products \cite{agilex_dsp}, and academic works proposing even lower precision multipliers in the Logic Array Blocks \cite{low_precision_soft_logic}, we will look to take advantage of our variable precision support to maintain accuracy and achieve even higher performance.

\section{Conclusion}
In this paper we have demonstrated that a gather-based layer-pipelined approach to CNN acceleration can leverage sparsity without the added hardware cost from prior works.  This approach enables inference throughput at a batch size of 1 that is 4x higher than the fastest GPU for machine learning on a large but sparse CNN.  On smaller and more efficient dense models that do not leverage our 0-weight skipping we still achieve higher throughput than the GPU and another FPGA accelerator while running at twice the precision.  Our variable precision support and throughput balancing algorithms will allow future accelerators based on this architecture to prune weights only from layers where accuracy does not suffer, and reduce the precision in particular layers where higher precision is less important.  Looking towards future FPGA architectures with DSP support for lower precision multipliers, these features could provide further performance improvements per area of 2x or more.

\Urlmuskip=0mu plus 1mu\relax
\bibliographystyle{IEEEtranN}
\bibliography{IEEEabrv,main}

\begin{thebibliography}{29}
\providecommand{\natexlab}[1]{#1}
\providecommand{\url}[1]{#1}
\csname url@samestyle\endcsname
\providecommand{\newblock}{\relax}
\providecommand{\bibinfo}[2]{#2}
\providecommand{\BIBentrySTDinterwordspacing}{\spaceskip=0pt\relax}
\providecommand{\BIBentryALTinterwordstretchfactor}{4}
\providecommand{\BIBentryALTinterwordspacing}{\spaceskip=\fontdimen2\font plus
\BIBentryALTinterwordstretchfactor\fontdimen3\font minus
  \fontdimen4\font\relax}
\providecommand{\BIBforeignlanguage}[2]{{%
\expandafter\ifx\csname l@#1\endcsname\relax
\typeout{** WARNING: IEEEtranN.bst: No hyphenation pattern has been}%
\typeout{** loaded for the language `#1'. Using the pattern for}%
\typeout{** the default language instead.}%
\else
\language=\csname l@#1\endcsname
\fi
#2}}
\providecommand{\BIBdecl}{\relax}
\BIBdecl

\bibitem[{Lu} et~al.(2019){Lu}, {Xie}, {Huang}, {Zhang}, {Lin}, and
  {Liang}]{efficient_sparse}
L.~{Lu}, J.~{Xie}, R.~{Huang}, J.~{Zhang}, W.~{Lin}, and Y.~{Liang}, ``An
  efficient hardware accelerator for sparse convolutional neural networks on
  fpgas,'' in \emph{2019 IEEE 27th Annual International Symposium on
  Field-Programmable Custom Computing Machines (FCCM)}, April 2019, pp. 17--25.

\bibitem[Molchanov et~al.(2016)Molchanov, Tyree, Karras, Aila, and
  Kautz]{pruning}
P.~Molchanov, S.~Tyree, T.~Karras, T.~Aila, and J.~Kautz, ``Pruning
  convolutional neural networks for resource efficient inference,'' in
  \emph{ICLR}, 2016.

\bibitem[Han et~al.(2016)Han, Liu, Mao, Pu, Pedram, Horowitz, and Dally]{eie}
\BIBentryALTinterwordspacing
S.~Han, X.~Liu, H.~Mao, J.~Pu, A.~Pedram, M.~A. Horowitz, and W.~J. Dally,
  ``Eie: Efficient inference engine on compressed deep neural network,'' in
  \emph{Proceedings of the 43rd International Symposium on Computer
  Architecture}, ser. ISCA '16.\hskip 1em plus 0.5em minus 0.4em\relax
  Piscataway, NJ, USA: IEEE Press, 2016, pp. 243--254. [Online]. Available:
  \url{https://doi.org/10.1109/ISCA.2016.30}
\BIBentrySTDinterwordspacing

\bibitem[Han et~al.(2017)Han, Kang, Mao, Hu, Li, Li, Xie, Luo, Yao, Wang, Yang,
  and Dally]{ese}
\BIBentryALTinterwordspacing
S.~Han, J.~Kang, H.~Mao, Y.~Hu, X.~Li, Y.~Li, D.~Xie, H.~Luo, S.~Yao, Y.~Wang,
  H.~Yang, and W.~B.~J. Dally, ``Ese: Efficient speech recognition engine with
  sparse lstm on fpga,'' in \emph{Proceedings of the 2017 ACM/SIGDA
  International Symposium on Field-Programmable Gate Arrays}, ser. FPGA
  '17.\hskip 1em plus 0.5em minus 0.4em\relax New York, NY, USA: ACM, 2017, pp.
  75--84. [Online]. Available: \url{http://doi.acm.org/10.1145/3020078.3021745}
\BIBentrySTDinterwordspacing

\bibitem[Zhang et~al.(2019)Zhang, Li, Wang, Liu, Qin, and Zhao]{sparse_fc}
M.~Zhang, L.~Li, H.~Wang, Y.~Liu, H.~Qin, and W.~Zhao, ``Optimized compression
  for implementing convolutional neural networks on fpga,'' \emph{Electronics},
  vol.~8, p. 295, 03 2019.

\bibitem[Cao et~al.(2019)Cao, Zhang, Yao, Xiao, Nie, Zhan, Liu, Wu, and
  Zhang]{bank_balanced}
\BIBentryALTinterwordspacing
S.~Cao, C.~Zhang, Z.~Yao, W.~Xiao, L.~Nie, D.~Zhan, Y.~Liu, M.~Wu, and
  L.~Zhang, ``Efficient and effective sparse lstm on fpga with bank-balanced
  sparsity,'' in \emph{Proceedings of the 2019 ACM/SIGDA International
  Symposium on Field-Programmable Gate Arrays}, ser. FPGA '19.\hskip 1em plus
  0.5em minus 0.4em\relax New York, NY, USA: ACM, 2019, pp. 63--72. [Online].
  Available: \url{http://doi.acm.org/10.1145/3289602.3293898}
\BIBentrySTDinterwordspacing

\bibitem[Kung et~al.(2019)Kung, McDanel, and Zhang]{packed_sparse}
\BIBentryALTinterwordspacing
H.~Kung, B.~McDanel, and S.~Q. Zhang, ``Packing sparse convolutional neural
  networks for efficient systolic array implementations: Column combining under
  joint optimization,'' in \emph{Proceedings of the Twenty-Fourth International
  Conference on Architectural Support for Programming Languages and Operating
  Systems}, ser. ASPLOS '19.\hskip 1em plus 0.5em minus 0.4em\relax New York,
  NY, USA: ACM, 2019, pp. 821--834. [Online]. Available:
  \url{http://doi.acm.org/10.1145/3297858.3304028}
\BIBentrySTDinterwordspacing

\bibitem[Zhang et~al.(2016)Zhang, Du, Zhang, Lan, Liu, Li, Guo, Chen, and
  Chen]{cambriconx}
\BIBentryALTinterwordspacing
S.~Zhang, Z.~Du, L.~Zhang, H.~Lan, S.~Liu, L.~Li, Q.~Guo, T.~Chen, and Y.~Chen,
  ``Cambricon-x: An accelerator for sparse neural networks,'' in \emph{The 49th
  Annual IEEE/ACM International Symposium on Microarchitecture}, ser.
  MICRO-49.\hskip 1em plus 0.5em minus 0.4em\relax Piscataway, NJ, USA: IEEE
  Press, 2016, pp. 20:1--20:12. [Online]. Available:
  \url{http://dl.acm.org/citation.cfm?id=3195638.3195662}
\BIBentrySTDinterwordspacing

\bibitem[Parashar et~al.(2017)Parashar, Rhu, Mukkara, Puglielli, Venkatesan,
  Khailany, Emer, Keckler, and Dally]{scnn}
\BIBentryALTinterwordspacing
A.~Parashar, M.~Rhu, A.~Mukkara, A.~Puglielli, R.~Venkatesan, B.~Khailany,
  J.~Emer, S.~W. Keckler, and W.~J. Dally, ``Scnn: An accelerator for
  compressed-sparse convolutional neural networks,'' in \emph{Proceedings of
  the 44th Annual International Symposium on Computer Architecture}, ser. ISCA
  '17.\hskip 1em plus 0.5em minus 0.4em\relax New York, NY, USA: ACM, 2017, pp.
  27--40. [Online]. Available: \url{http://doi.acm.org/10.1145/3079856.3080254}
\BIBentrySTDinterwordspacing

\bibitem[Aimar et~al.(2017)Aimar, Mostafa, Calabrese, Rios-Navarro,
  Tapiador-Morales, Lungu, Milde, Corradi, Linares-Barranco, Liu, and
  Delbruck]{nullhop}
A.~Aimar, H.~M.~Z. Mostafa, E.~Calabrese, A.~Rios-Navarro, R.~Tapiador-Morales,
  I.-A. Lungu, M.~B. Milde, F.~Corradi, A.~Linares-Barranco, S.-C. Liu, and
  T.~Delbruck, ``Nullhop: A flexible convolutional neural network accelerator
  based on sparse representations of feature maps,'' \emph{IEEE Transactions on
  Neural Networks and Learning Systems}, vol.~30, pp. 644--656, 2017.

\bibitem[{Lavin} and {Gray}(2016)]{original_winograd}
A.~{Lavin} and S.~{Gray}, ``Fast algorithms for convolutional neural
  networks,'' in \emph{2016 IEEE Conference on Computer Vision and Pattern
  Recognition (CVPR)}, June 2016, pp. 4013--4021.

\bibitem[Aydonat et~al.(2017)Aydonat, O'Connell, Capalija, Ling, and
  Chiu]{original_dla}
\BIBentryALTinterwordspacing
U.~Aydonat, S.~O'Connell, D.~Capalija, A.~C. Ling, and G.~R. Chiu, ``An
  opencl\texttrademark deep learning accelerator on arria 10,'' in
  \emph{Proceedings of the 2017 ACM/SIGDA International Symposium on
  Field-Programmable Gate Arrays}, ser. FPGA '17.\hskip 1em plus 0.5em minus
  0.4em\relax New York, NY, USA: ACM, 2017, pp. 55--64. [Online]. Available:
  \url{http://doi.acm.org/10.1145/3020078.3021738}
\BIBentrySTDinterwordspacing

\bibitem[Lu and Liang(2018)]{sparse_winograd}
\BIBentryALTinterwordspacing
L.~Lu and Y.~Liang, ``Spwa: An efficient sparse winograd convolutional neural
  networks accelerator on fpgas,'' in \emph{Proceedings of the 55th Annual
  Design Automation Conference}, ser. DAC '18.\hskip 1em plus 0.5em minus
  0.4em\relax New York, NY, USA: ACM, 2018, pp. 135:1--135:6. [Online].
  Available: \url{http://doi.acm.org/10.1145/3195970.3196120}
\BIBentrySTDinterwordspacing

\bibitem[Vincent et~al.(2017)Vincent, Stephano, Frumkin, Ginsburg, and
  Demouth]{numerical_stability_winograd}
K.~Vincent, K.~Stephano, M.~A. Frumkin, B.~Ginsburg, and J.~Demouth, ``On
  improving the numerical stability of winograd convolutions,'' in \emph{ICLR},
  2017.

\bibitem[Zhang and Prasanna(2017)]{fft_cnn}
\BIBentryALTinterwordspacing
C.~Zhang and V.~Prasanna, ``Frequency domain acceleration of convolutional
  neural networks on cpu-fpga shared memory system,'' in \emph{Proceedings of
  the 2017 ACM/SIGDA International Symposium on Field-Programmable Gate
  Arrays}, ser. FPGA '17.\hskip 1em plus 0.5em minus 0.4em\relax New York, NY,
  USA: ACM, 2017, pp. 35--44. [Online]. Available:
  \url{http://doi.acm.org/10.1145/3020078.3021727}
\BIBentrySTDinterwordspacing

\bibitem[Tan and Le(2019)]{efficientnet}
\BIBentryALTinterwordspacing
M.~Tan and Q.~Le, ``{E}fficient{N}et: Rethinking model scaling for
  convolutional neural networks,'' in \emph{Proceedings of the 36th
  International Conference on Machine Learning}, ser. Proceedings of Machine
  Learning Research, K.~Chaudhuri and R.~Salakhutdinov, Eds., vol.~97.\hskip
  1em plus 0.5em minus 0.4em\relax Long Beach, California, USA: PMLR, 09--15
  Jun 2019, pp. 6105--6114. [Online]. Available:
  \url{http://proceedings.mlr.press/v97/tan19a.html}
\BIBentrySTDinterwordspacing

\bibitem[{Fowers} et~al.(2018){Fowers}, {Ovtcharov}, {Papamichael},
  {Massengill}, {Liu}, {Lo}, {Alkalay}, {Haselman}, {Adams}, {Ghandi}, {Heil},
  {Patel}, {Sapek}, {Weisz}, {Woods}, {Lanka}, {Reinhardt}, {Caulfield},
  {Chung}, and {Burger}]{brainwave}
J.~{Fowers}, K.~{Ovtcharov}, M.~{Papamichael}, T.~{Massengill}, M.~{Liu},
  D.~{Lo}, S.~{Alkalay}, M.~{Haselman}, L.~{Adams}, M.~{Ghandi}, S.~{Heil},
  P.~{Patel}, A.~{Sapek}, G.~{Weisz}, L.~{Woods}, S.~{Lanka}, S.~K.
  {Reinhardt}, A.~M. {Caulfield}, E.~S. {Chung}, and D.~{Burger}, ``A
  configurable cloud-scale dnn processor for real-time ai,'' in \emph{2018
  ACM/IEEE 45th Annual International Symposium on Computer Architecture
  (ISCA)}, June 2018, pp. 1--14.

\bibitem[{Moreau} et~al.(2019){Moreau}, {Chen}, {Vega}, {Roesch}, {Yan},
  {Zheng}, {Fromm}, {Jiang}, {Ceze}, {Guestrin}, and {Krishnamurthy}]{vta}
T.~{Moreau}, T.~{Chen}, L.~{Vega}, J.~{Roesch}, E.~{Yan}, L.~{Zheng},
  J.~{Fromm}, Z.~{Jiang}, L.~{Ceze}, C.~{Guestrin}, and A.~{Krishnamurthy}, ``A
  hardware–software blueprint for flexible deep learning specialization,''
  \emph{IEEE Micro}, vol.~39, no.~5, pp. 8--16, Sep. 2019.

\bibitem[Noronha et~al.(2018)Noronha, Salehpour, and Wilton]{legup}
\BIBentryALTinterwordspacing
D.~H. Noronha, B.~Salehpour, and S.~J.~E. Wilton, ``Leflow: Enabling flexible
  {FPGA} high-level synthesis of tensorflow deep neural networks,''
  \emph{CoRR}, vol. abs/1807.05317, 2018. [Online]. Available:
  \url{http://arxiv.org/abs/1807.05317}
\BIBentrySTDinterwordspacing

\bibitem[Sharma et~al.(2016)Sharma, Park, Mahajan, Amaro, Kim, Shao, Mishra,
  and Esmaeilzadeh]{dnnweaver}
H.~Sharma, J.~Park, D.~Mahajan, E.~Amaro, J.~K. Kim, C.~Shao, A.~Mishra, and
  H.~Esmaeilzadeh, ``From high-level deep neural models to fpgas,'' in
  \emph{Microarchitecture (MICRO), 2016 49th Annual IEEE/ACM International
  Symposium on}.\hskip 1em plus 0.5em minus 0.4em\relax IEEE, 2016, pp. 1--12.

\bibitem[Chen et~al.(2019)Chen, He, Zhang, Hao, and Chen]{cloud_dnn}
Y.~Chen, J.~He, X.~Zhang, C.~Hao, and D.~Chen, ``Cloud-dnn: An open framework
  for mapping dnn models to cloud fpgas,'' 02 2019, pp. 73--82.

\bibitem[Mishra et~al.(2018)Mishra, Nurvitadhi, Cook, and Marr]{wrpn}
\BIBentryALTinterwordspacing
A.~Mishra, E.~Nurvitadhi, J.~J. Cook, and D.~Marr, ``{WRPN}: Wide
  reduced-precision networks,'' in \emph{International Conference on Learning
  Representations}, 2018. [Online]. Available:
  \url{https://openreview.net/forum?id=B1ZvaaeAZ}
\BIBentrySTDinterwordspacing

\bibitem[Rybalkin et~al.(2018)Rybalkin, Pappalardo, Ghaffar, Gambardella, Wehn,
  and Blott]{finn_l}
V.~Rybalkin, A.~Pappalardo, M.~M. Ghaffar, G.~Gambardella, N.~Wehn, and
  M.~Blott, ``Finn-l: Library extensions and design trade-off analysis for
  variable precision lstm networks on fpgas,'' \emph{2018 28th International
  Conference on Field Programmable Logic and Applications (FPL)}, pp. 89--897,
  2018.

\bibitem[Deng et~al.(2009)Deng, Dong, Socher, Li, Li, and Fei-Fei]{imagenet}
J.~Deng, W.~Dong, R.~Socher, L.-J. Li, K.~Li, and L.~Fei-Fei, ``{ImageNet: A
  Large-Scale Hierarchical Image Database},'' in \emph{CVPR09}, 2009.

\bibitem[{NVIDIA Corporation}(2019)]{v100}
\BIBentryALTinterwordspacing
{NVIDIA Corporation}. (2019) Nvidia tesla deep learning product performance.
  [Online]. Available:
  \url{https://web.archive.org/web/20190817114937/https://developer.nvidia.com/deep-learning-performance-training-inference}
\BIBentrySTDinterwordspacing

\bibitem[He et~al.(2016)He, Zhang, Ren, and Sun]{resnet}
K.~He, X.~Zhang, S.~Ren, and J.~Sun, ``Deep residual learning for image
  recognition,'' 06 2016, pp. 770--778.

\bibitem[{Wu} et~al.(2019){Wu}, {Zhang}, {Jia}, {Tian}, {Li}, {Sui}, {Xie}, and
  {Shan}]{xilinx_layer_pipe}
D.~{Wu}, Y.~{Zhang}, X.~{Jia}, L.~{Tian}, T.~{Li}, L.~{Sui}, D.~{Xie}, and
  Y.~{Shan}, ``A high-performance cnn processor based on fpga for mobilenets,''
  in \emph{2019 29th International Conference on Field Programmable Logic and
  Applications (FPL)}, Sep. 2019, pp. 136--143.

\bibitem[{Intel Corporation}(2019)]{agilex_dsp}
{Intel Corporation}. (2019) Intel agilex variable precision dsp blocks user
  guide.

\bibitem[Boutros et~al.(2019)Boutros, Eldafrawy, Yazdanshenas, and
  Betz]{low_precision_soft_logic}
\BIBentryALTinterwordspacing
A.~Boutros, M.~Eldafrawy, S.~Yazdanshenas, and V.~Betz, ``Math doesn't have to
  be hard: Logic block architectures to enhance low-precision
  multiply-accumulate on fpgas,'' in \emph{Proceedings of the 2019 ACM/SIGDA
  International Symposium on Field-Programmable Gate Arrays}, ser. FPGA
  '19.\hskip 1em plus 0.5em minus 0.4em\relax New York, NY, USA: ACM, 2019, pp.
  94--103. [Online]. Available:
  \url{http://doi.acm.org/10.1145/3289602.3293912}
\BIBentrySTDinterwordspacing

\end{thebibliography}


\end{document}